\documentclass[11pt]{article}
\usepackage[utf8]{inputenc}
\pdfoutput=1
\usepackage{amssymb,amsmath,mathrsfs,enumerate}
\usepackage{mathtools}
\usepackage{graphicx,rotate,multicol}
\usepackage{float}
\usepackage{tocloft}
\usepackage{subfig}
\usepackage[margin=10pt,labelfont=bf]{caption}
\usepackage{cite}
\usepackage{soul}
\usepackage{xcolor}
\usepackage[colorlinks=true,
            linkcolor=blue,
            urlcolor=blue,
            citecolor=teal]{hyperref}
\usepackage{multirow}
\usepackage{array}
 \usepackage{placeins}

\makeatletter
\let\Hy@linktoc\Hy@linktoc@page
\makeatother

\usepackage{color}
\definecolor{ourcolor}{rgb}{0.7, 0.25, 0.05}
\usepackage{tikz,braket}
\long\def\rpl#1!!#2!!{\textcolor{red}{#1} \textcolor{blue}{#2}}

\let\bar=\overline

\def \order(#1){{\mathcal O} \left(#1 \right)}

\evensidemargin=\oddsidemargin
\allowdisplaybreaks

\title{\color{black}{\bf Neutrinos from captured dark matter annihilation in a galactic population of neutron stars}}

\author {{\bf Debajit Bose,} $^{a}$ \footnote{debajitbose550@gmail.com}
\hspace{4pt}  {\bf Tarak Nath Maity,} $^{b}$  \footnote{tarak.maity.physics@gmail.com} \hspace{4pt}  { \bf and Tirtha Sankar Ray} $^{a}$ \footnote{tirthasankar.ray@gmail.com}
\\[10pt]
$^{a}$\,{\it Department of Physics, Indian Institute of Technology Kharagpur, Kharagpur 721302, India} \\
$^{b}$\,{\it Centre  for  High  Energy  Physics,  Indian  Institute  of  Science,  Bangalore  560012,  India}
}

\date{}

\begin{document}

\maketitle

\begin{abstract}
Particulate dark matter captured by a population of neutron stars distributed around the galactic center while annihilating through long-lived mediators can give rise to an observable neutrino flux. We examine the prospect of an idealised gigaton detector like IceCube/KM3NeT in probing such scenarios. Within this framework, we report an improved reach in spin-dependent and spin-independent dark matter nucleon cross-section below the current limits for dark matter masses in the TeV-PeV range.   
\end{abstract}

\newpage

\hrule \hrule
\tableofcontents
\vskip 10pt
\hrule \hrule 

\section{Introduction}
\label{sec:intro}

Particulate dark matter (DM) that can annihilate to Standard Model (SM) species in the present day Universe may be observed through indirect detection experiments \cite{Slatyer:2017sev}. The impediment to such detection is the low generic density of DM. This makes the patches of the galaxy where the DM may accumulate particularly interesting. In this context, considerable attention has been recently focused on the capture of DM in celestial bodies like the Sun, exoplanets, neutron stars, etc. that can locally boost the DM population. In this paper, we discuss the possibility of DM captured in galactic neutron star (NS) population and its subsequent annihilation to neutrinos resulting in observable signals in the Earth based detectors. 

The idea of DM capture is quite simple, as the astrophysical object roams around the galaxy, it wades through a sea of DM particles. Consequently, gravitationally focused DM particles can scatter off the stellar constituents depositing its energy partially. After the scattering, if the final velocity of the DM particle becomes less than the escape velocity it gets trapped \cite{Press:1985ug,Gould:1987ju,Gould:1987ir,Gould1992BigBA,Gould:1999je}. The celestial object can heat up by absorbing the annihilation products and the kinetic energy of the trapped DM. The heating signatures of captured DM has been well studied in the literature \cite{Kouvaris:2007ay, Goldman:1989nd,Baryakhtar:2017dbj,Raj:2017wrv,Bell:2018pkk,Garani:2018kkd,Dasgupta:2019juq,Bell:2019pyc, Camargo:2019wou, Joglekar:2019vzy,Acevedo:2019agu,Bell:2020jou,Joglekar:2020liw,Bramante:2017xlb,Ilie:2020vec, Dasgupta:2020dik,Garani:2020wge,Bell:2020lmm,Leane:2020wob,Bell:2020obw,Bell:2021fye,NathMaity:2021cne,McKeen:2021jbh,Ilie:2021umw,Anzuini:2021lnv}. The possibility of such capture mechanism leading to the formation of black holes has been pointed out in \cite{McDermott:2011jp, deLavallaz:2010wp, Kouvaris:2010jy, Busoni:2013kaa, Bell:2013xk, Dasgupta:2020mqg}.

An alternate strategy would be the search for annihilation products of the captured DM. While direct annihilation of captured DM to SM states leads to heating of the object, if the annihilation takes place via a long-lived mediator \cite{Pospelov:2007mp,Pospelov:2008jd,Batell:2009zp,Dedes:2009bk,Fortes:2015qka,Okawa:2016wrr,Yamamoto:2017ypv,Holdom:1985ag,Holdom:1986eq,Chen:2009ab,Rothstein:2009pm,Berlin:2016gtr,Cirelli:2016rnw,Cirelli:2018iax}, the mediator can easily escape that object and its subsequent decay to SM states can be detected on the Earth. Neutrinos and photons generated from captured DM annihilation can act as detectable messengers for the Earth based experiments \cite{Silk:1985ax, Krauss:1985aaa, Jungman:1994jr, Pospelov:2008jd, Batell:2009zp,Bell:2011sn,IceCube:2012ugg, Baratella:2013fya, Danninger:2014xza,Smolinsky:2017fvb,Leane:2017vag,HAWC:2018szf,Nisa:2019mpb,Niblaeus:2019gjk,Xu:2020lrn,Bell:2021pyy,Leane:2021tjj,Xu:2021glr,Bell:2021esh,Tonnis:2021krs, Fornengo:2017lax, Super-Kamiokande:2015xms}.

Due to their high density, neutron stars are ideal candidates to probe really low DM-nucleon cross-section through DM capture. The galactic center has a dense stellar environment which is optimal for DM capture due to the high density of both stars and DM. Gamma-ray signal from the captured DM annihilation in the population of neutron stars near the galactic center has been studied recently in \cite{Leane:2021ihh}. In this paper, we concentrate on the neutrinos produced from the annihilation of captured DM in the galactic center population of neutron stars. We examine the prospects of gigaton neutrino detectors like IceCube \cite{IceCube:2016yoy} or upcoming KM3NeT \cite{KM3Net:2016zxf} in detecting these neutrinos. We find that these neutrino telescopes can probe both spin-dependent (SD) and spin-independent (SI) DM-nucleon scattering cross-section orders of magnitude below the current limits for DM masses in the range $\mathcal{O}(10)$ TeV - $\mathcal{O}(10)$ PeV.

The rest of this paper is organised as follows. In section \ref{sec:capture}, we briefly review the capture of DM in the distribution of neutron stars near the galactic center. In section \ref{sec:nuflux}, we calculate the neutrino flux arising from the annihilation of captured DM 
in NS distribution and determine the sensitivity of the neutrino detectors in section \ref{sec:sensitivity}, before we conclude in section \ref{sec:conclusion}.

\section{Dark matter capture in galactic neutron stars}
\label{sec:capture}

In this section, we briefly review the mechanism of DM capture by celestial bodies. As an astrophysical object roams through the galactic medium, it encounters numerous ambient DM particles. These DM particles fall into the steep gravitational potential of the celestial object. The velocity of the DM particle at the surface is given by $w = \sqrt{u^2 + v_{\rm esc}^2}$, where $u$ is the ambient DM velocity and $v_{\rm esc}$ is the escape velocity of the celestial object. These gravitationally focused DM particles interact with the stellar constituents and lose their energy. If the final velocity of the DM particle becomes less than the escape velocity, it is captured within the gravitational potential of the celestial body. 

Within the multi-scatter capture framework \cite{Bramante:2017xlb}, the probability that DM particle scatters $S$ times within the celestial object is given by 
\begin{equation}
\label{pN}
p_S(\tau) = 2 \int_0^1 dy \, \frac{y \, e^{-y \tau} (y \tau)^S}{S!},
\end{equation}
where the optical depth, $\tau = 1.5 \, \sigma_{\chi n}/ \sigma_{\rm sat}$, $\sigma_{\chi n}$ is the DM nucleon cross-section and $\sigma_{\rm sat}$ is the saturation cross-section, defined as $\sigma_{\rm sat} = \pi \, R^2/N_n$. Here $N_n$ is the number of nucleons within the astrophysical object. The total capture rate within a celestial body situated at a distance $r$ from the galactic center is given by \cite{Bramante:2017xlb},
\begin{equation}
\label{CN}
C(r) = \sum_{S=1}^{\infty} \frac{\pi \, R^2 \, p_S(\tau)}{(1-\frac{2 \, G \, M}{R})} \frac{\sqrt{6}}{3 \sqrt{\pi} \, \bar{v}} \frac{\rho_{\chi}(r)}{m_{\chi}} \left( (2 \, \bar{v}^2 + 3 \, v_{\rm esc}^2) - (2 \, \bar{v}^2 + 3 \, v_{S}^2) \, \, {\rm exp} \left[- \frac{3 (v_S^2 - v_{\rm esc}^2)}{2 \, \bar{v}^2} \right] \right),
\end{equation}

where $M$, $R$ being the mass, radius of the concerned celestial body respectively, $G$ is the Gravitational constant, $\bar{v}$ is the DM velocity dispersion which is $\sim 100 \, {\rm km/s}$ in our region of interest, as can be seen from \cite{Sofue2013RotationCA,Sofue2016RotationAM,Sofue2020RotationCO}. To remain conservative, we have set the value to the solar neighborhood one at $\bar{v} \approx 220 \, {\rm km/s}$. In Eq.\,(\ref{CN}), $\rho_{\chi} (r)$ is the density of DM in the neighbourhood of the celestial object and $m_{\chi}$ is the mass of the DM. The term $v_S$ is defined as $v_S = v_{\rm esc} (1-\frac{\beta_{+}}{2})^{- S/2}$ with $\beta_{+} = 4 \, m_{\chi} \, m_n / (m_{\chi} + m_n)^2$ which is related to the DM energy loss by the scattering event.  As it is evident from Eq.\,\eqref{pN}, the scattering probability of large number of scatterings is highly suppressed making it sufficient to truncate the summation in Eq.\,\eqref{CN} for a finite value of $S$ which is larger than the optical depth. We have incorporated the leading relativistic corrections to the capture rate.

There is compelling evidence that the galactic center can host a considerable population of neutron stars \cite{Freitag:2006qf,Hopman:2006xn,Merritt_2010,Generozov:2018niv}. Recent observations of O/B stars near the galactic center indicate an enhanced rate of \emph{in situ} neutron star and black hole formation \cite{Genzel:2003cn,Levin:2003kp}. The radial profile of neutron stars near the galactic center is estimated considering both the historical injection and the continuous formation of the \emph{in situ} population. Due to its high density of neutron stars, the galactic center is an attractive region of the sky to probe for DM capture. The generalization of the DM capture due to this population of NS is straightforward and can be written as
\begin{equation}
\label{captotal}
C_{\rm tot} = 4 \, \pi \int_{r_1}^{r_2} r^2 \, n_{\rm NS} (r) \: C(r) \: dr,
\end{equation}
where $n_{\rm NS} (r)$ is the neutron star density function adopted from \cite{Generozov:2018niv}. We will consider a typical neutron star mass at ${\rm 1.5 \, M_\odot}$ with a radius of $10 \, {\rm km}$. We integrate over the region $r = 0.1 - 10 \, {\rm pc}$, beyond which the NS density sharply falls off. As can be seen from Eq.\,\eqref{captotal}, along with the distribution of neutron stars, total capture also depend on the DM density profile around the galactic center. In this work, we have utilized both cored and cuspy spherically-symmetric DM density profiles to calculate the total capture rate. The considered density profiles are given in Table \ref{tab:profiles}. The cuspy density profiles predict higher concentration of DM particles near the galactic center as compared to cored DM profiles. Expectedly the total DM capture rate for the cuspy profiles are higher than the cored ones.

\begin{table}[t] 
\centering
\begin{tabular}{|| l | l | m{3.7cm} ||}
\hline
Profile Name & Radial Density Function & Parameters \\
\hline \hline
NFW \cite{Navarro:1995iw,Navarro:1996gj} & $ \rho_{\rm NFW}(r) = \frac{\rho_0}{\left( \frac{r}{r_s} \right) \left[ 1+ \left(\frac{r}{r_s} \right)\right]^2} $ & $r_s = 20 \, {\rm kpc}$ \cite{Fornasa:2013iaa,Gaskins:2016cha} \\
\hline
Moore \cite{Moore:1999gc} & $ \rho_{\rm Moore}(r) = \frac{\rho_0}{\left( \frac{r}{r_s} \right)^{\gamma} \left[1 + \left( \frac{r}{r_s} \right) \right]^{3 - \gamma}} $ & $r_s = 12 \, {\rm kpc}$ \cite{Leane:2021ihh}; \newline $\gamma = 1.5$ \cite{Leane:2020wob,Leane:2021ihh,Gaskins:2016cha,Moore:1999gc} \\
\hline
Einsato \cite{1965TrAlm...5...87E} & $ \rho_{\rm Ein}(r) = \rho_0 \, {\rm exp} \left[ - \left( \frac{2}{\alpha} \right) \left\{ \left( \frac{r}{r_s} \right)^{\alpha} - 1 \right\} \right] $ & $r_s = 20 \, {\rm kpc}$ \cite{Fornasa:2013iaa,Gaskins:2016cha}; \newline $\alpha = 0.2$ \cite{Fornasa:2013iaa,Gaskins:2016cha,Iocco:2011jz} \\
\hline
Burkert \cite{Burkert:1995yz} & $ \rho_{\rm Bur}(r) = \frac{\rho_0}{\left( 1+ \frac{r}{r_s} \right) \left[ 1+ \left( \frac{r}{r_s} \right)^2 \right]} $ & $r_s = 6 \, {\rm kpc}$ \cite{Fornasa:2013iaa,Gaskins:2016cha} \\
\hline
\end{tabular}
\caption{Various DM density profiles and the corresponding fit parameters for the Milky Way galaxy. The distributions are normalised to have a local DM density at $0.3 {\rm \, GeV / cm^3}$.}
\label{tab:profiles}
\end{table}

\section{Neutrino flux from captured DM annihilation}
\label{sec:nuflux}

In the previous section, we have discussed the formalism of DM capture in the population of neutron stars which is distributed around the galactic center. The time variation of the number density of trapped DM inside a neutron star, $N_{\chi}$ is given by
\begin{equation}
\label{dNdt}
\frac{d N_{\chi}}{d t} = C_{\rm tot} - C_{\rm ann} \, N_{\chi}^2 - C_{\rm evap} \, N_{\chi},
\end{equation}
where $C_{\rm tot}$ is the total capture rate of DM, $C_{\rm ann}$ is the DM annihilation coefficient and $C_{\rm evap}$ is the evaporation rate of DM. For the DM mass range of interest we can safely neglect the effect of $C_{\rm evap}$ \cite{Garani:2021feo}. Once equilibrium is reached, the annihilation rate of the captured DM can be read as
\begin{equation}
\label{Gammaann}
\Gamma_{\rm ann} = \frac{1}{2} \, C_{\rm ann} \, N_{\chi}^2 = \frac{1}{2} \, C_{\rm tot}
\end{equation}

The annihilation rate inside neutron stars depends on the thermalization and equilibration of captured DM. These assumptions do not hold for small DM-nucleon cross-sections leading to a significant suppression in the annihilation rate. For a conservative neutron star lifetime of 10 Myr, we find that the captured DM can thermalize \cite{Bertoni:2013bsa} and equilibrate \cite{Bramante:2017xlb,Kouvaris:2010vv} in our region of interest. If the DM particles directly annihilate to SM states, they get trapped inside the star leading to its heating. However, in a broad class of scenarios the DM may annihilate to SM states through long-lived mediators which can escape the core of the neutron star \cite{Pospelov:2007mp,Pospelov:2008jd,Batell:2009zp,Dedes:2009bk,Fortes:2015qka,Okawa:2016wrr,Yamamoto:2017ypv,Holdom:1985ag,Holdom:1986eq,Chen:2009ab,Rothstein:2009pm,Berlin:2016gtr,Cirelli:2016rnw,Cirelli:2018iax}. Subsequently, the decay product of the mediator may provide observable signals in neutrino and $\gamma-$ray telescopes. However, there is a possibility that the mediator may scatter with the SM constituents within the neutron star before emerging that results in considerable attenuation of the obtained flux. As demonstrated in \cite{Leane:2021ihh}, for reasonable choice of model parameters attenuation can be significantly reduced. Hence, we assume that all of the mediator particles eventually decay outside the neutron stars and produce detectable SM flux. Captured DM annihilating through long-lived mediator have been previously explored in the context of the Sun \cite{Silk:1985ax, Krauss:1985aaa, Jungman:1994jr, Pospelov:2008jd, Batell:2009zp,Bell:2011sn,IceCube:2012ugg, Baratella:2013fya, Danninger:2014xza,Smolinsky:2017fvb,Leane:2017vag,HAWC:2018szf,Nisa:2019mpb,Niblaeus:2019gjk,Xu:2020lrn,Bell:2021pyy,Leane:2021tjj,Xu:2021glr,Bell:2021esh} and recently in the context of other celestial bodies \cite{Leane:2021tjj,Leane:2021ihh}. We consider the DM capture by a galactic distribution of neutron stars annihilating through sufficiently long-lived mediator. We focus on the impact of these annihilation products on the Earth based neutrino experiments.

\begin{figure}
\begin{center}
\includegraphics[scale=0.22]{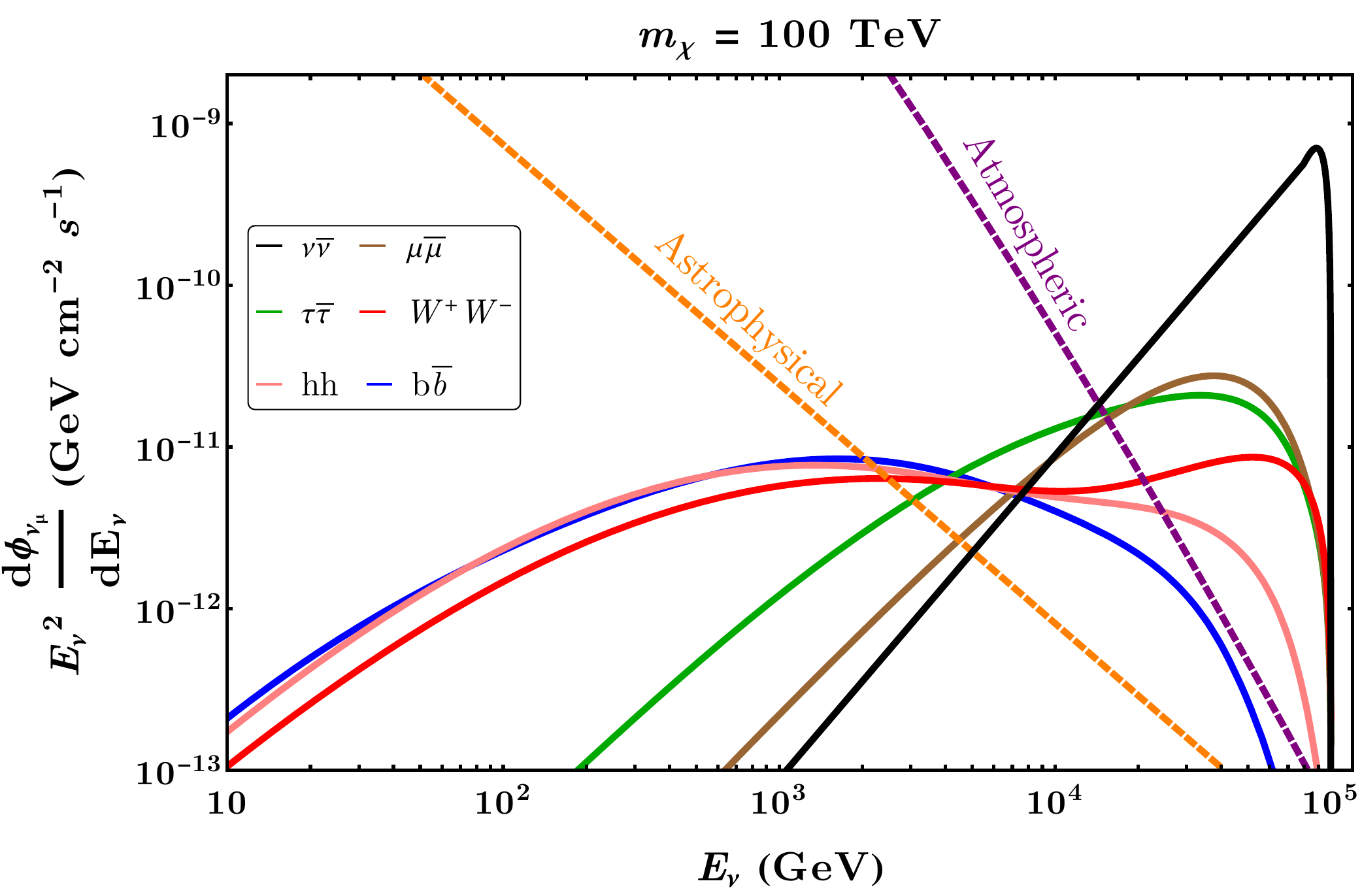}
\caption{Muon neutrino flux at the surface of the Earth for various SM final states : $\tau$ (green), $b$-${\rm quark}$ (blue), $\nu$ (black), $\mu$ (brown), ${\rm Higgs}$ (pink) and $W$-${\rm boson}$ (red) with DM mass, $m_{\chi} = 100 \, {\rm TeV}$, mediator mass, $m_Y = 2 \, {\rm TeV}$ and $\sigma_{\chi n} = 10^{-45} \, {\rm cm^2}$ for Moore DM profile. The background atmospheric (purple) and astrophysical (orange) muon neutrino fluxes are shown with dashed lines.}
\label{fig:flux}
\end{center}
\end{figure}

Considering the Earth based neutrino detectors are optimized for detecting muon neutrinos we concentrate on the muon neutrino flux originating from DM capture discussed in section \ref{sec:capture}. The differential muon neutrino flux reaching the Earth from the captured DM annihilation through the long-lived mediator is given by
\begin{equation}
E_{\nu}^2 \, \frac{d \phi_{\nu_{\mu}}}{d E_{\nu}} = \frac{\Gamma_{\rm ann}}{4 \, \pi \, D^2} \times {\rm Br}(Y \rightarrow {\rm SM} \, \bar{\rm SM}) \times \left( \frac{1}{3} E_{\nu}^2 \frac{d N_{\nu}}{d E_{\nu}} \right)  \times  \left( e^{-\frac{R}{\eta c \tau_{Y}}} - e^{-\frac{D}{\eta c \tau_{Y}}} \right),
\label{eq:flux}
\end{equation}
where $D$ is the distance of the Earth from the NS population, ${\rm Br}(Y \rightarrow {\rm SM} \, \bar{\rm SM})$ is the branching ratio of mediator ($Y$) decay to SM states, $dN_{\nu}/d E_{\nu}$ is the neutrino spectrum. We have estimated the total neutrino spectrum at production due to all neutrino flavors \cite{Elor:2015bho}. As we are dealing with neutrinos coming from distances of thousands of parsecs, we can safely neglect the oscillation effects and assume the flavor ratio of neutrinos reaching the Earth is $\nu_e:\nu_{\mu}:\nu_{\tau}=1:1:1$. In presenting our results, we have assumed $100 \, \%$ branching ratio to individual SM final states. The last term in the parenthesis in Eq.\,\eqref{eq:flux} represents the survival probability of neutrino to reach the Earth. We will consider the mediator to be long-lived. This can be achieved by considering either sufficiently long lifetime ($\tau_Y$) of the mediator or sufficiently large boost, $\eta = m_{\chi}/m_Y$. These conditions can lead to the decay length of the mediator $L \approx \eta c \tau_Y > R$. It is assumed that all the neutrinos produced by the mediator traverses the distance $\sim \, 8.3 \, {\rm kpc}$ to reach the detectors on the Earth. Given these assumptions, the neutrino flux in Eq.\,\eqref{eq:flux} is uniquely determined by DM mass, mediator mass and DM-nucleon cross-section. The muon neutrino flux for DM mass $m_{\chi}=100$ TeV, mediator mass $m_Y=2$ TeV and $\sigma_{\chi n} = 10^{-45} \, {\rm cm^{2}}$ is displayed in Fig.\,\ref{fig:flux} for various different SM final state channels for Moore DM density profile defined in section \ref{sec:capture}.

In our framework it is assumed that the captured DM dominantly annihilate through the long-lived mediators. Interestingly, the DM capture can be driven by an independent DM-nucleon interaction which contributes sub-dominantly to the annihilation process. Even with this kind of setup, equilibrium between the capture and annihilation can be achieved simply because of the hierarchy of the nucleon and captured DM density inside a neutron star. Remaining agnostic to specific DM models our framework can be premised on a large class of scenarios including secluded DM model \cite{Pospelov:2007mp,Pospelov:2008jd,Batell:2009zp,Dedes:2009bk,Fortes:2015qka,Okawa:2016wrr,Yamamoto:2017ypv}, dark photon model \cite{Holdom:1985ag,Holdom:1986eq}, dark higgs model \cite{Chen:2009ab}, etc. Moreover, for long-lived mediator the unitarity bound \cite{Griest:1989wd} may be evaded by considering late injection of entropy \cite{Berlin:2016gtr,Cirelli:2016rnw, Cirelli:2018iax}.

\section{Detection prospects at gigaton $\nu$ detectors}
\label{sec:sensitivity}

In this section, we will discuss the reach of a km$^3$ neutrino telescope in detecting the neutrino flux considered in the previous section.  In the detectors the incoming neutrino flux is detected via the corresponding muon produced through charge-current (CC) interactions generating track like events \cite{Kistler:2006hp}. The other possibility is cascade events produced by $\nu_e$ and $\nu_{\tau}$ through both CC and neutral-current (NC) interactions \cite{Kistler:2006hp}. At present relatively weaker angular resolution of the cascade events makes them difficult to detect and will not be considered here.

The number of contained muons produced per unit energy for a given muon neutrino flux is given by \cite{Kistler:2006hp}
\begin{equation}
\label{startmuon}
\frac{d N_{\mu}}{d E_{\mu}} \simeq N_A \, \rho \, V \, T \, \frac{1}{1-y} \left[ \frac{d \phi_{\nu_{\mu}}}{d E_{\nu}}(E_{\nu}) \, \sigma_{\rm CC}(E_{\nu})\right]_{E_{\nu}=\frac{E_{\mu}}{1-y}},
\end{equation}
where $N_A$ is the Avogadro number, $\rho \simeq 1 \, {\rm g/cm^3}$ is the density of the detector material. $V$ and $T$ are the effective volume and effective exposure time of the detector respectively. The cross-section of neutrino CC interaction ($\sigma_{\rm CC}$) is obtained by averaging over $\nu-$N and $\bar{\nu}-$N interactions extracted from \cite{Gandhi:1995tf}. Note that average muon energy $\left\langle E_{\mu} \right\rangle$ can be related to neutrino energy $E_{\nu}$, by $\left\langle E_{\mu} \right\rangle = E_{\mu} = E_{\nu} (1-y)$. The parameter $y$ which determines the fraction of neutrino energy that has been transferred to muons is kept fixed at 0.4 \cite{Gandhi:1995tf,Leane:2017vag}.

The major source of background for this neutrino search is the atmospheric neutrino background. We have taken all sky averaged muon neutrino flux from \cite{Honda:2015fha} and extrapolate it by using the parametric form given in \cite{Sinegovskaya:2014pia}. Another possible source for such high energetic neutrinos is the astrophysical neutrinos emanating from galactic and extra-galactic sources. While there is a considerable uncertainty in its estimation, we adopt the distribution from \cite{Stettner:2019tok}. Note that for both the components, background fluxes are estimated by considering $\nu-\mu$ opening angle, $\theta_{\nu \mu} = 0.7^{\circ} (\rm TeV/E_{\nu})^{0.6}$ \cite{KM3NeT2008}. The background muon neutrino flux from the atmospheric and astrophysical sources are summarized in Fig.\,\ref{fig:flux}.

Throughout our analysis we have not considered the atmospheric muon background owing to the fact that atmospheric muon can be reduced significantly by using veto at the detectors. For KM3NeT, this can be done using the time when the galactic center remains below the horizon which is $\sim 37\%$ of the total exposure time \cite{Ng:2020ghe}. For this exposure the Earth can shield the large atmospheric muon background. Note that for this exposure neutrinos will have to traverse within the Earth leading to an attenuation. The attenuation is significant at large energies which can be a factor $\sim 2$ in neutrino flux at $\sim 1$ PeV energy, as reported in \cite{Ng:2020ghe}. To remain conservative we have multiplied the obtained neutrino flux by a factor of half at all energy scales to obtain our limits. For IceCube, the galactic center remains above the horizon throughout. The events from the galactic center are observed as down-going muon events at the detector which might submerge in the atmospheric muon background. Hence to detect the events from the galactic center at IceCube one can utilize a smaller inner volume within the detector so that the external regions can act as muon shield \cite{ANTARES:2020leh}. Admittedly the KM3NeT is more optimised to observe these neutrino signals from the galactic center.

\begin{figure*}[t]
\begin{center}
\subfloat[\label{sf:silimit}]{\includegraphics[scale=0.21]{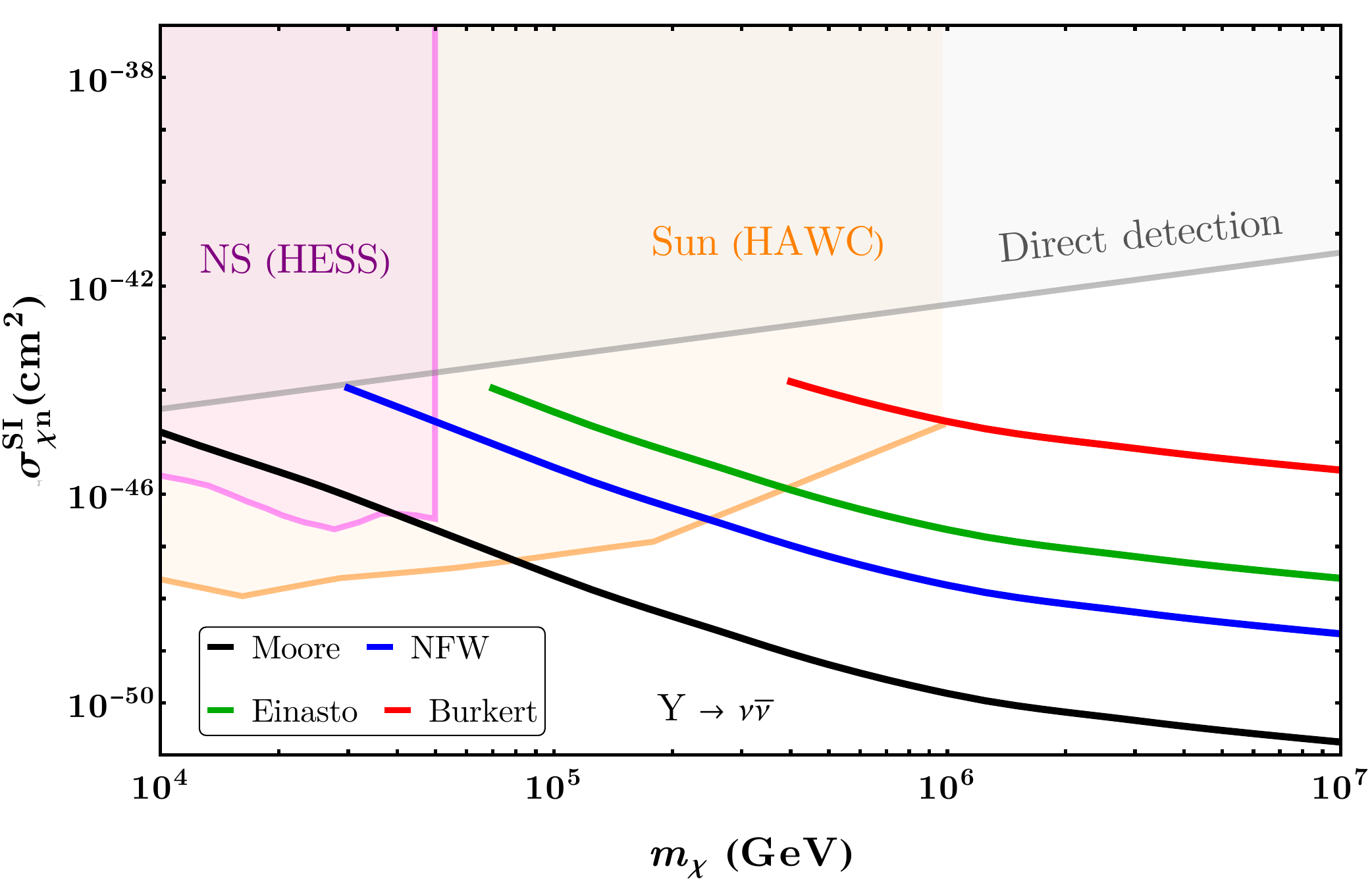}}~~~
\subfloat[\label{sf:sdlimit}]{\includegraphics[scale=0.21]{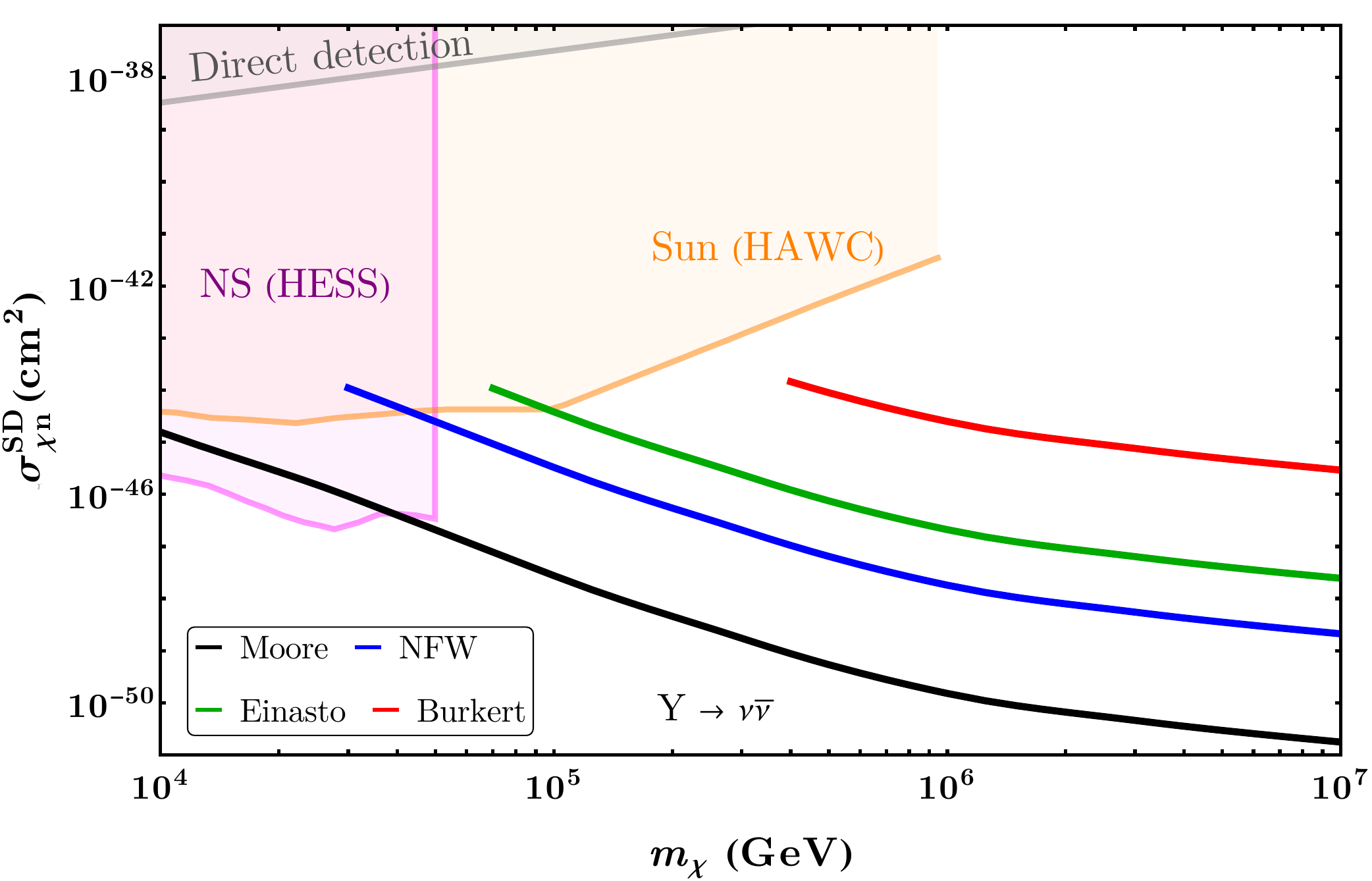}}
\caption{Sensitivities of SI (left) and SD (right) DM-nucleon scattering cross-sections for neutrino channel due to captured DM annihilation in a galactic NS population for the  DM density profiles given in Table\,\ref{tab:profiles} with a mediator mass of $2 \, {\rm TeV}$. We have shown the direct detection limits  \cite{PICO:2019vsc,PandaX-4T:2021bab} and also the constraints from captured DM annihilation to photons for the Sun \cite{HAWC:2018szf,Bell:2021pyy}, NS population \cite{Leane:2021ihh}.}
\label{fig:limits}
\end{center}
\end{figure*}

To estimate the sensitivity, we have calculated the number of signal events and the total background events in individual energy bins [${\rm max}(E_{\rm thres},m_{\chi}/5),m_{\chi}$] following \cite{Mandal:2009yk,Dasgupta:2012bd}. The energy binning is carefully chosen to be greater than the energy resolution of a typical gigaton detector \cite{Drakopoulou:2016azl, IceCube:2010whx, KM3Net:2016zxf}. Our limits are conservative by requiring that signal events to be as high as the background events to be detected. In Fig.\,\ref{fig:limits}, we show the SI and SD limits for neutrino decay channel of the mediator of mass $m_Y = 2 \, {\rm TeV}$. We have also shown the variation in the limits for different DM profiles discussed in section \ref{sec:capture}. Additionally the bounds are sensitive to the decay channel of the mediator. Though in Fig.\,\ref{fig:limits}, we have shown the most optimistic limits for neutrino channel, we have summarised the sensitivities due to all the other prominent channels in appendix \ref{app:Otherlimits}. The limits are determined by the interplay of the capture rate and the background neutrino flux presented in Fig.\,\ref{fig:flux}. As the background event rate decreases with energy, the sensitivity increases with DM mass. There is a distinctive change in slope of the exclusion limits at $m_{\chi} \sim 10^6$\,GeV. This can be traced back to the combined effect of the change in the scaling of the capture rate with $m_{\chi}$ and the domination of the astrophysical neutrino background around the PeV scale. We have plotted the extrapolated SD direct detection limits using PICO-60  results \cite{PICO:2019vsc} and recently reported PandaX-4T results for SI interactions \cite{PandaX-4T:2021bab}, which at present is more constraining than the XENON1T limits \cite{XENON:2018voc}.
We have also shown indirect detection limits due to captured DM annihilation to $\gamma$-rays and subsequent probes at HAWC and HESS telescopes for the Sun \cite{HAWC:2018szf,Bell:2021pyy} and NS population \cite{Leane:2021ihh} respectively. While for neutron stars, the obtained SI and SD limits are similar, for the Sun, the limits for SI cross-section are orders of magnitude stringent due to enhanced capture rate. We check that the cross-section of interest is consistent with the equilibrium limit of the neutron stars that remain below the region depicted in Fig.\,\ref{fig:limits}. This makes the galactic distribution of neutron stars a promising source to probe smaller DM-nucleon cross-section.

\section{Conclusions}
\label{sec:conclusion}

While all the pieces of evidence for DM are gravitational, it is phenomenologically interesting to consider a non-gravitational interaction between SM and DM and raise the possibility of detection on the Earth based experiments. Capture of DM by astrophysical objects due to DM-nucleon interaction in the core of the stars are of recent interest. The captured DM species can trigger observable changes in the properties of the host celestial objects that can provide on handle to probe DM indirectly. In this paper, we have explored a complimentary scenario where DM accumulated through capture in galactic neutron star population, annihilate to the sufficiently long-lived mediator. This long-lived mediator may escape the neutron star and decay to different SM states to produce an observable neutrino flux.

To study the feasibility of probing this DM induced neutrino flux, we have considered an idealised gigaton detector. In the detector, neutrinos can be detected mainly through muon tracks produced from the charged current interaction.  To obtain the limits on the parameter space, we have adopted a conservative approach of requiring the signal event to be as high as the leading background from atmospheric and astrophysical neutrinos. We find that the neutrinos from accumulated DM in a galactic population of neutron stars can constrain the spin-dependent and spin-independent cross-sections of DM orders of magnitude below the existing bounds for TeV-PeV scale DM.

\paragraph*{Acknowledgments\,:} We thank Nicole Bell for valuable comments. We thank Debanjan Bose and Ranjan Laha  for useful discussions. DB acknowledges MHRD, Government of India for fellowship. TNM thanks IOE-IISc fellowship program for financial assistance.

\vspace{- 0.2 cm}

\appendix
\section{Limits for other decay channels of the mediator}
\label{app:Otherlimits}

\begin{figure}[t]
\begin{center}
	\includegraphics[angle=0.0,width=0.39\textwidth]{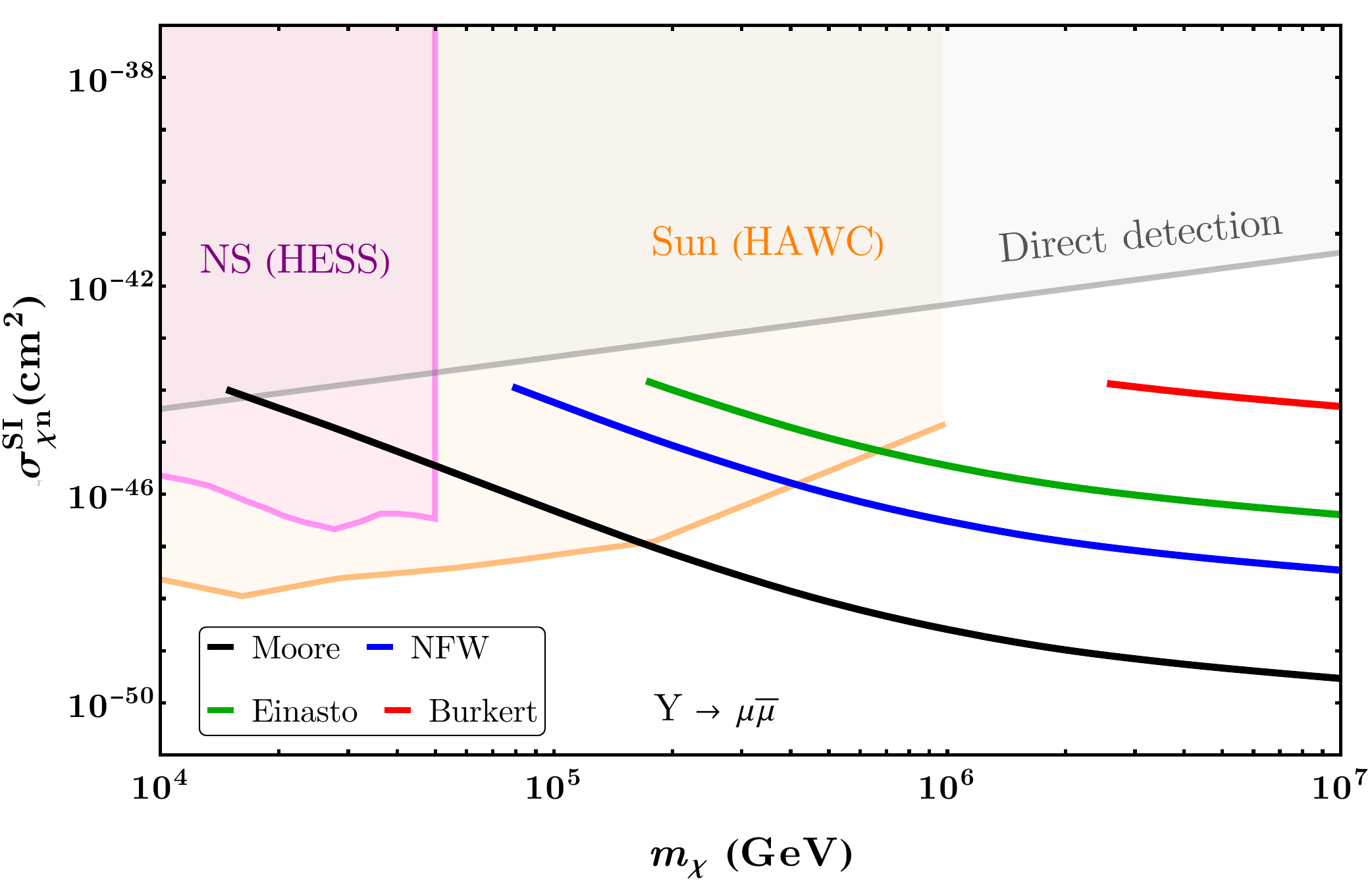}	~~
	\includegraphics[angle=0.0,width=0.39\textwidth]{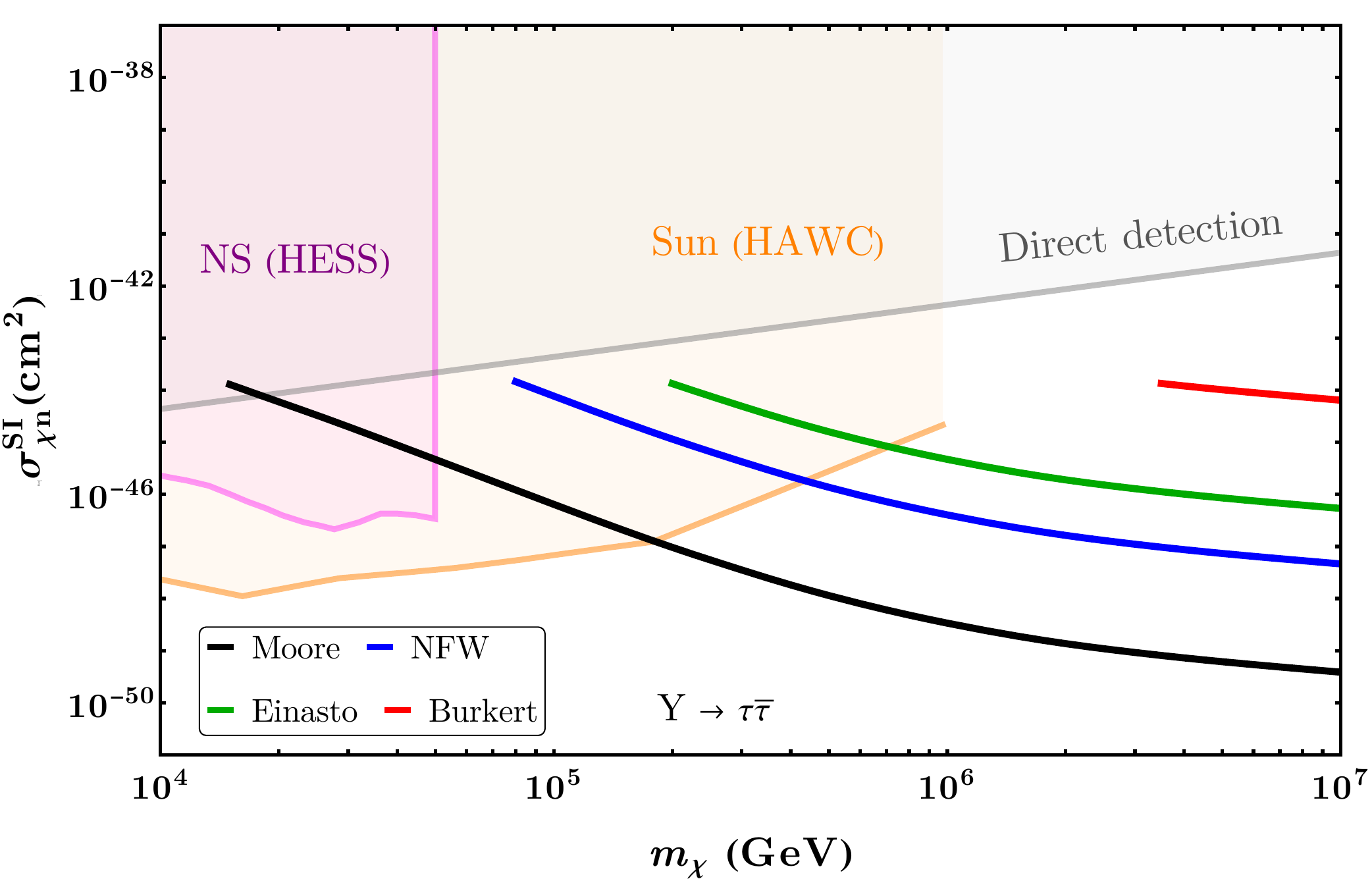} ~~ \\
	\includegraphics[angle=0.0,width=0.39\textwidth]{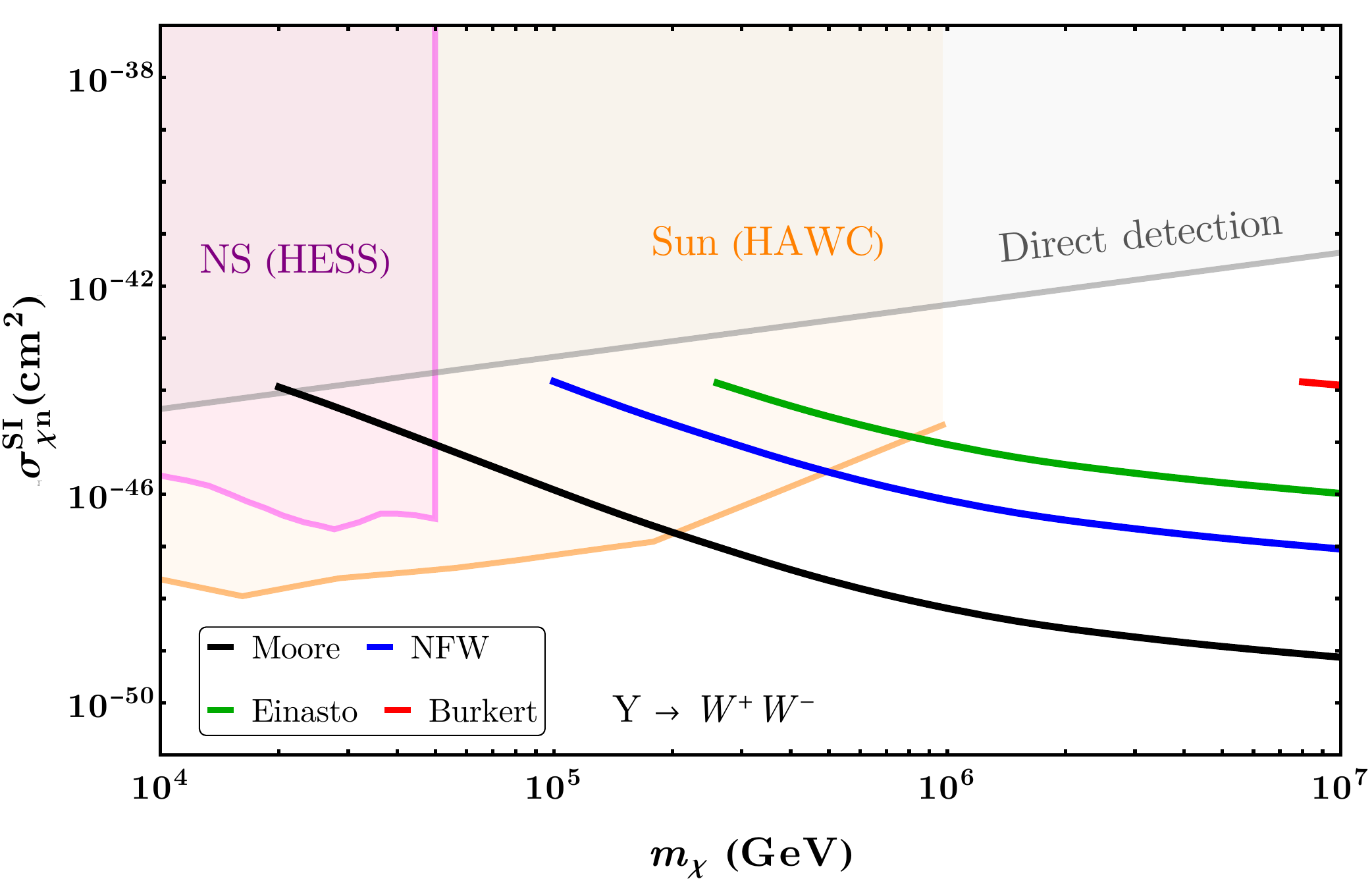} ~~
	\includegraphics[angle=0.0,width=0.39\textwidth]{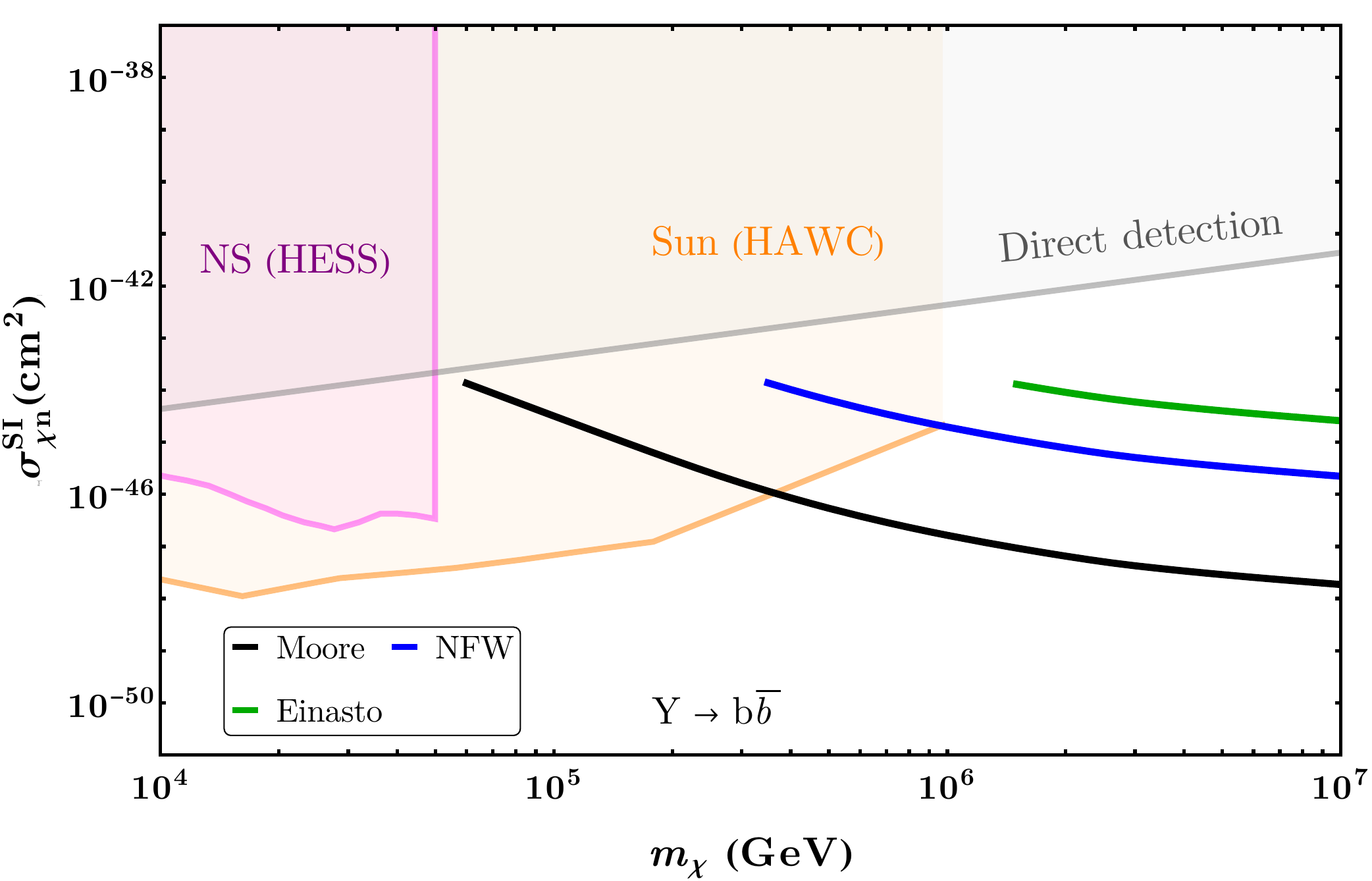} ~~	
	\caption{Exclusion limits of SI DM-nucleon scattering cross-section for various decay channels of the mediator of mass $2 \, {\rm TeV}$ for the  DM density profiles given in Table\,\ref{tab:profiles}. Direct detection limit \cite{PandaX-4T:2021bab} and other indirect search sensitivities for the Sun \cite{Bell:2021pyy} and neutron stars \cite{Leane:2021ihh} are also presented.}
\label{fig:other_si_limits}
\end{center}
\end{figure}

\begin{figure}[H]
\begin{center}
	\includegraphics[angle=0.0,width=0.39\textwidth]{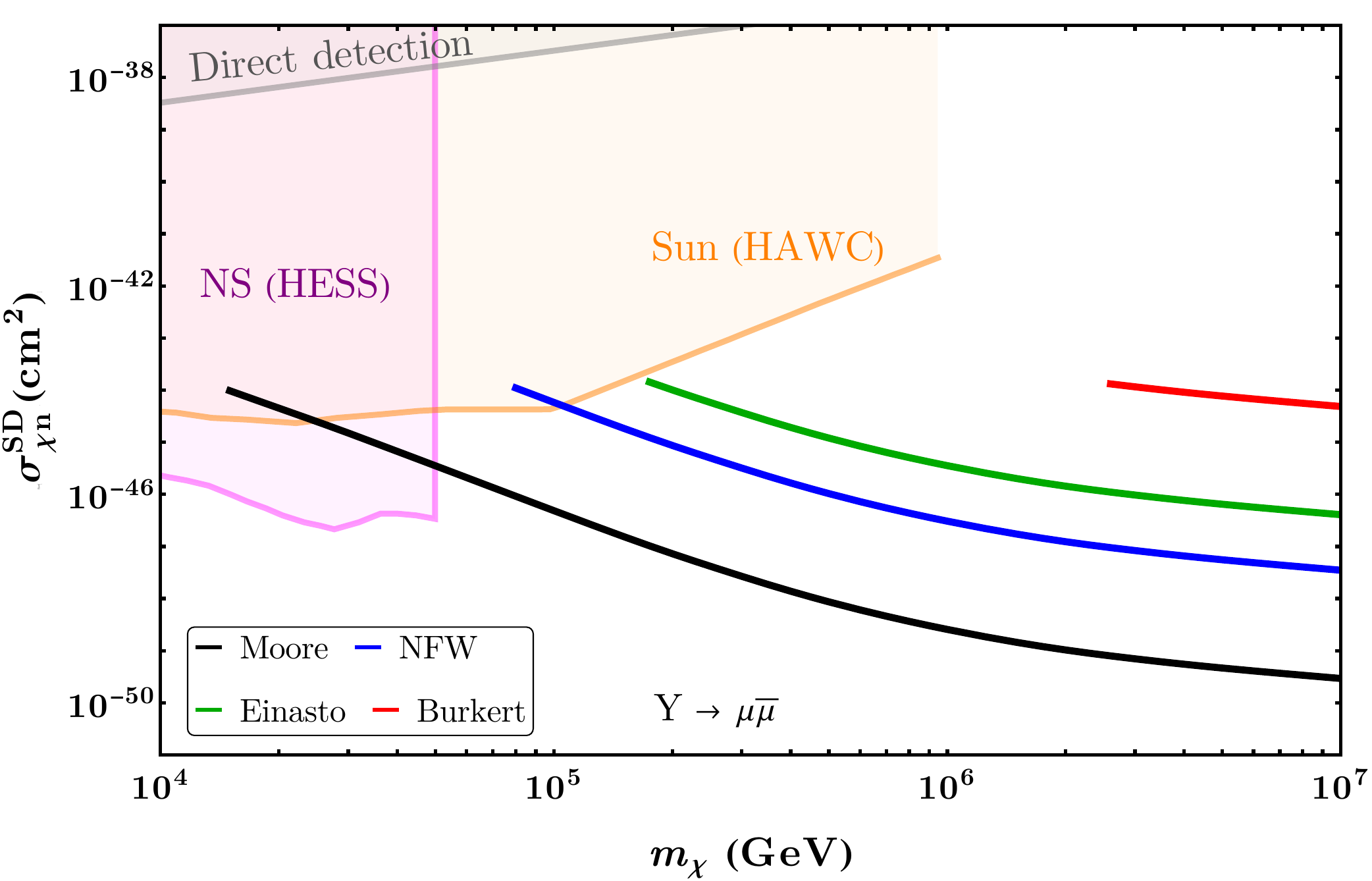}	~~
	\includegraphics[angle=0.0,width=0.39\textwidth]{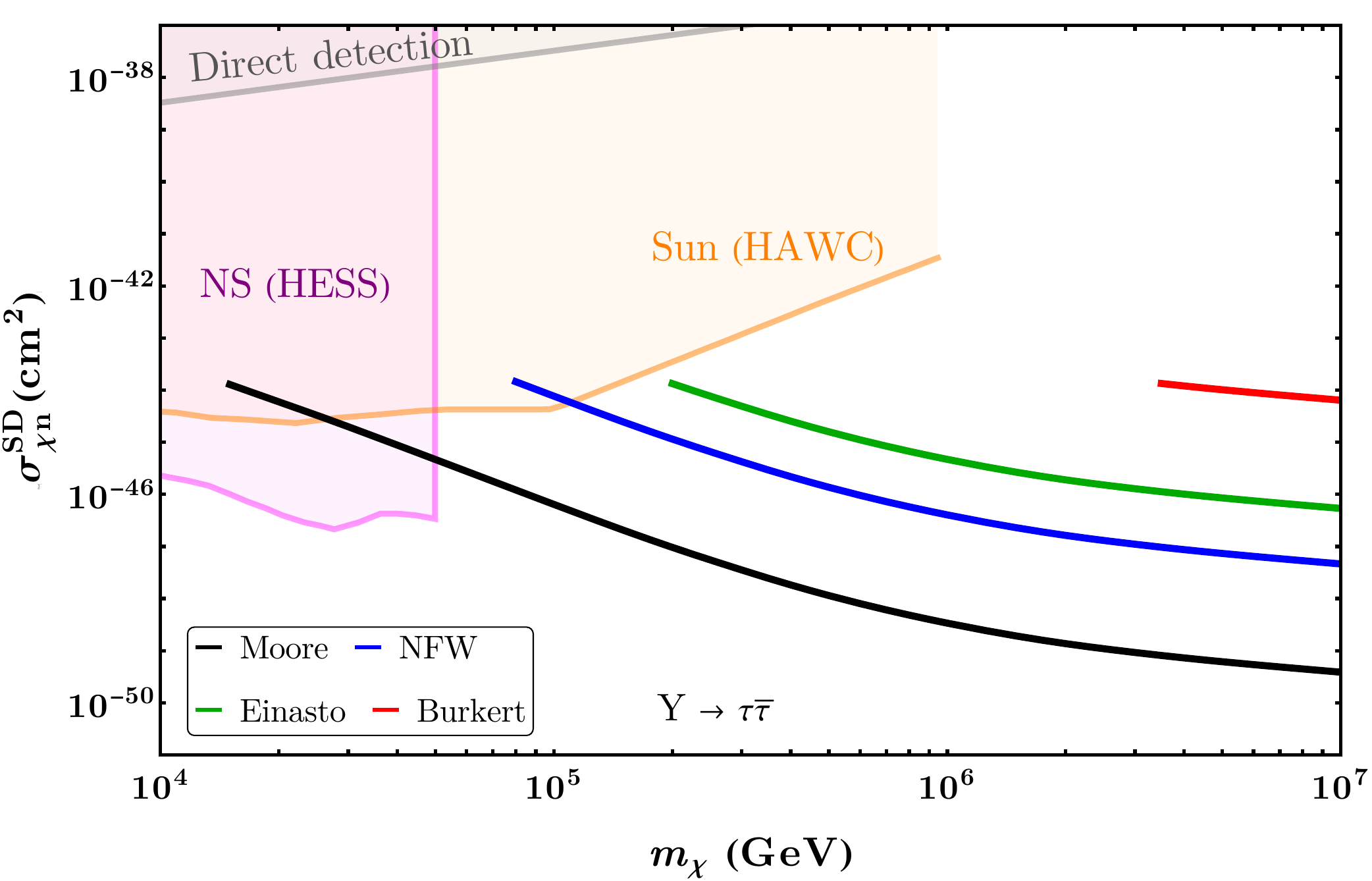} ~~ \\
	\includegraphics[angle=0.0,width=0.39\textwidth]{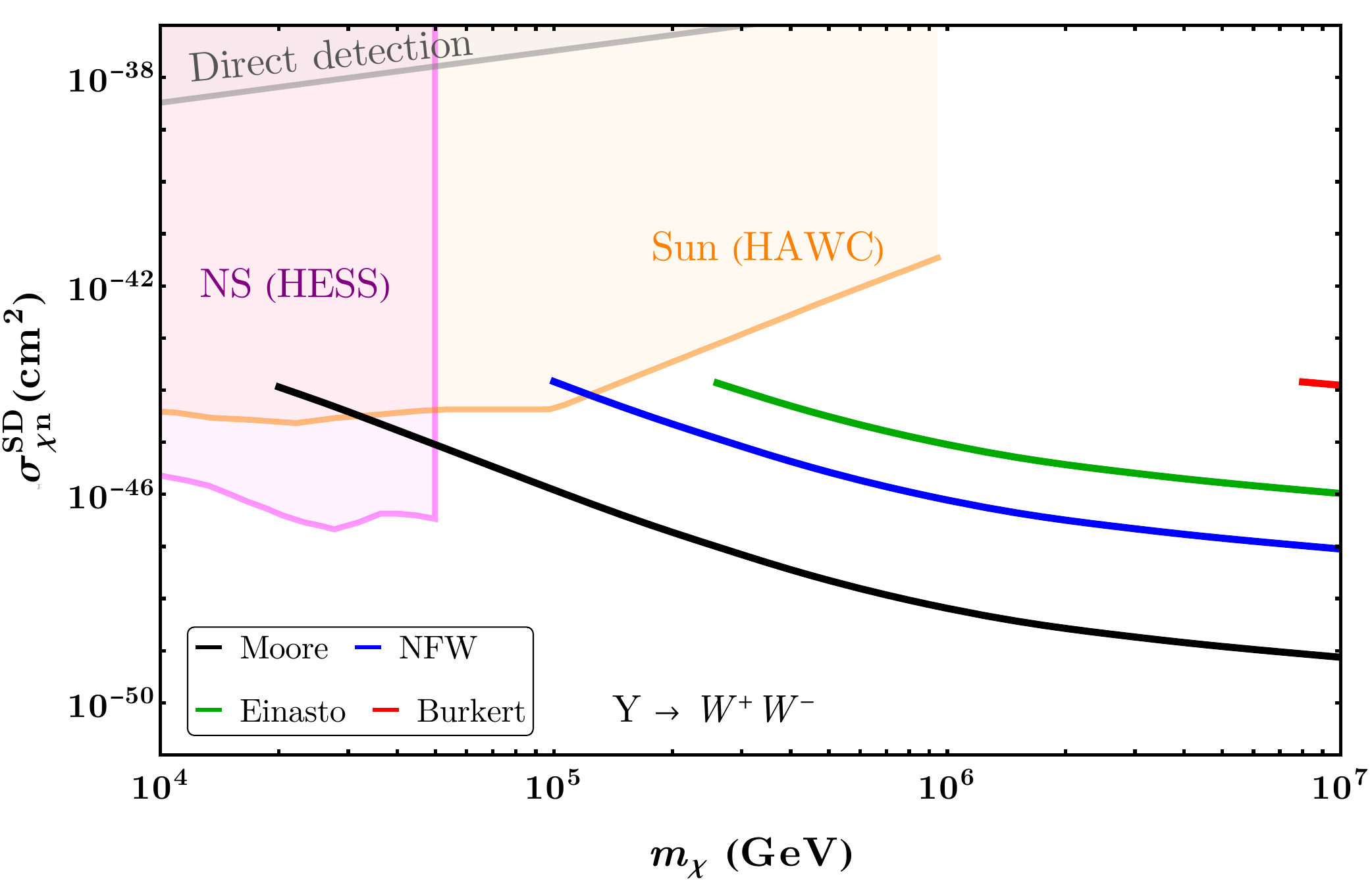} ~~
	\includegraphics[angle=0.0,width=0.39\textwidth]{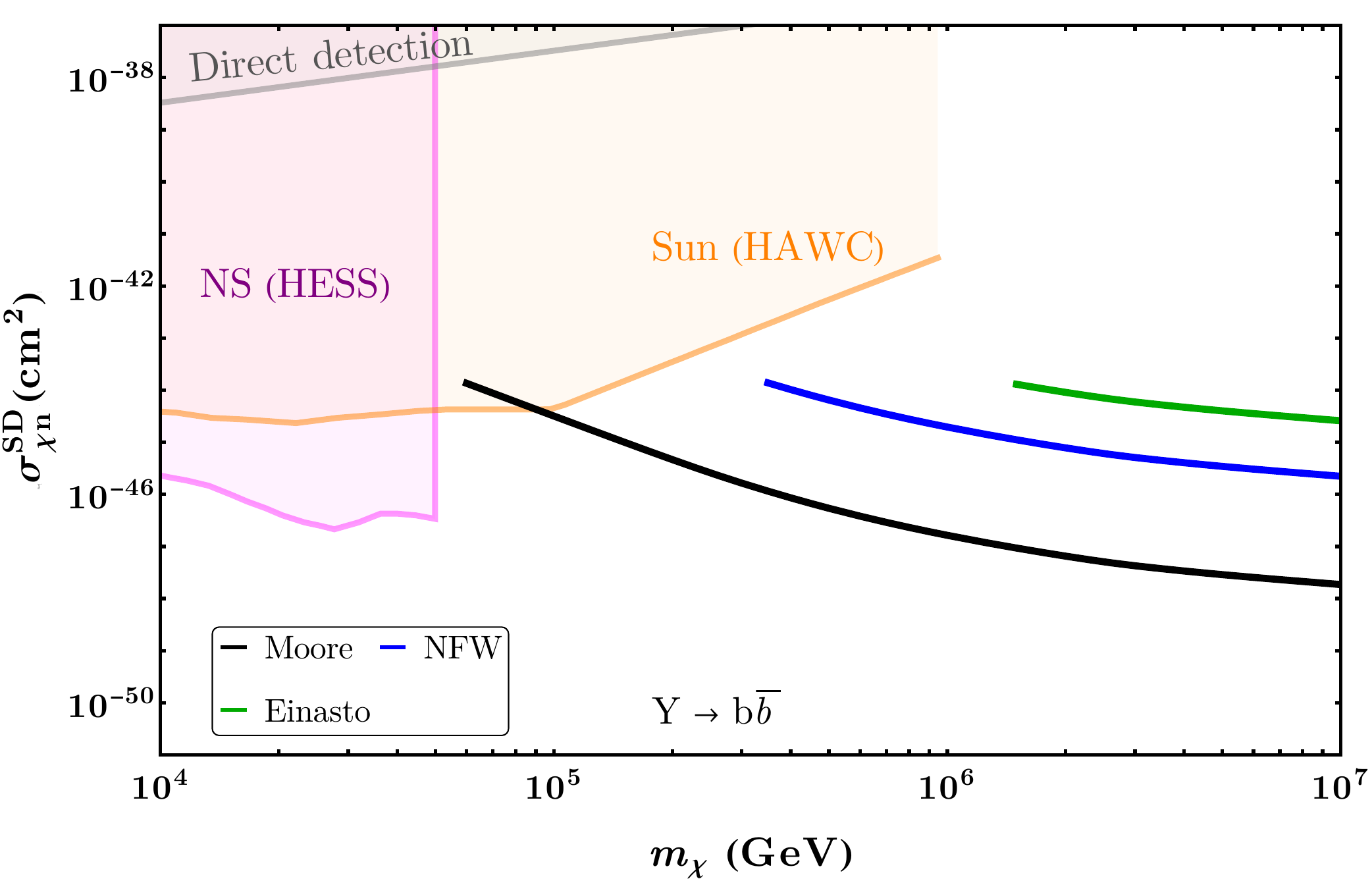} ~~	
	\caption{Same as Fig.\,\ref{fig:other_si_limits} but for SD DM-nucleon scattering cross-section. Direct detection limit \cite{PICO:2019vsc} and other indirect search sensitivities for the Sun \cite{HAWC:2018szf,Bell:2021pyy} and neutron stars \cite{Leane:2021ihh} are also presented.}
\label{fig:other_sd_limits}
\end{center}
\end{figure}

In this appendix, we will present the limits for some prominent decay channels of the long-lived mediator. The differential muon neutrino flux for different SM final state channels is shown in Fig.\,\ref{fig:flux}. The spectra for direct neutrino decay is peaked around the DM mass. The neutrino spectrum for other decay channels is relatively soft compared to direct neutrino channel. As the softness of the spectrum increases the obtained limits becomes weaker. We have demonstrated this in Figs.\,\ref{fig:other_si_limits} and \ref{fig:other_sd_limits} for SI and SD DM-nucleon cross section respectively. For some of the SM final state channels, there are no limits for cored Burkert profile which predicts relatively lower DM density near the galactic center.

\bibliographystyle{JHEP}
\bibliography{ref.bib}

\end{document}